\documentclass[pre,floatfix,superscriptaddress,twocolumn,showpacs]{revtex4-1}
\usepackage{graphicx}

\usepackage[utf8]{inputenc}
\usepackage[T1]{fontenc}
\usepackage{graphicx}
\usepackage{amsmath,amssymb}
\usepackage{hyperref}
\usepackage{nameref}
\usepackage{xcolor}
\hypersetup{
    colorlinks=true,
    linkcolor=blue,
    citecolor=blue,
    urlcolor=blue
}
\usepackage{tensor}
\usepackage{textgreek}
\usepackage{bm}
\usepackage{mathrsfs}

\usepackage{xcolor} 

\usepackage{multibib}

\makeatletter
\renewcommand{\@fnsymbol}[1]{%
  \ifcase#1\or * \or †\or ‡\or **\or \ddagger\or \mathsection\or \mathparagraph\or \|\or **\else\@arabic{#1}\fi}
\makeatother

\begin{document}
\title{Curvature-induced host-mediated polarization of active particles}
\author{Giulia Janzen}
\email{gjanzen@ucm.es}
\affiliation{Department of Structure of Matter, Thermal Physics, and Electronics, Complutense University of Madrid, Madrid, 28040, Spain}
\author{D. A. Matoz-Fernandez}
\email{dmatoz@ucm.es}
\affiliation{Department of Theoretical Physics, Complutense University of Madrid, Madrid, 28040, Spain}

\date{\today}

\begin{abstract}
Polar collective motion commonly arises from alignment interactions, particle anisotropy, or an imposed directional bias. Here we identify a distinct route to polar order that does not rely on alignment interactions between the active particles. We show that non-aligning active Brownian particles embedded in a dense passive medium can develop polar coherence when confined to a compact curved surface. Persistent active motion redistributes stress through the host and creates passive-depleted regions. When the stress-spreading length becomes comparable to the sphere radius, these regions merge into elongated scars that channel active motion and, through feedback with the active flux, promote a common direction of motion. Removing the passive host suppresses polar coherence even though the active particles continue to cluster on the same sphere. Our results establish an environment-mediated route to collective polarity in which symmetry breaking emerges from the coupling between active motion, passive stress redistribution, and compact geometry.
\end{abstract}

\keywords{active matter, curved surfaces, active-passive mixtures, mechanical coupling, polar order}

\maketitle

Collective motion in active matter usually requires an explicit source of orientational symmetry breaking. In polar flocks, long-range order arises through alignment interactions, as in Vicsek-type models and Toner--Tu hydrodynamics~\cite{Vicsek1995,Toner1998} and in natural flocks~\cite{Ballerini2008,Cavagna2010}. A related route is provided by self-aligning polar active matter, where collective motion arises from a phenomenological coupling between particle polarity and velocity rather than from explicit mutual alignment~\cite{Baconnier2025}. In active nematics and anisotropic active particles, order is organized by defect dynamics, particle shape, directed propulsion, or anisotropic interactions~\cite{Sanchez2012,DeCamp2015,Shankar2019,Baskaran2008,Bar2020,Riedel2024}. These works have revealed how active systems organize once a symmetry-breaking ingredient is present. What remains unclear is whether a non-polar active system can develop collective polarity through the environment in which it moves.

Curvature provides one way for the environment to act on active matter. Curved interfaces are common in living and synthetic systems, from epithelial monolayers and organoids to active nematics, bacterial films and colloids on droplets or vesicles~\cite{Luciano2021,Tang2022,Brandstatter2023,Keber2014,Ellis2018,Iyer2023,Nemeth2025}. Spherical confinement of active matter has also been realized experimentally in vesicle-based systems, making the sphere a directly accessible setting for studying how compactness and topology shape collective dynamics~\cite{Keber2014,Hsu2022}. More generally, curved surfaces and confining boundaries can redirect trajectories, focus transport, reorganize topological defects, and constrain ordered phases~\cite{Sknepnek2015,Shankar2017,Schonhofer2022,Mackay2026,Guruciaga2026}. Other environmental features, including obstacles, confinement, and passive components, can also reshape active motion by modifying the space or medium through which particles move~\cite{Weber2016, Codina2022}. Here we identify a different role for curvature. Rather than only guiding active trajectories or frustrating pre-existing order, curvature changes how stress is transmitted through a passive host. This allows the surrounding material to supply the symmetry-breaking field needed to polarize otherwise non-aligning dopants.

We study non-aligning active particles embedded in a dense passive medium and confined to a spherical surface. At large persistence, the active particles develop coherent polar motion, although neither the particles nor the host contains an imposed polar axis. This order emerges from the interplay between geodesic motion and host-mediated interactions. Together, curvature and the passive host create local directional biases that correlate active trajectories and generate polar streams without an explicit alignment rule. We call this mechanism scar-mediated polarization.

\begin{figure*}[t]
\centering
\includegraphics[width=\textwidth]{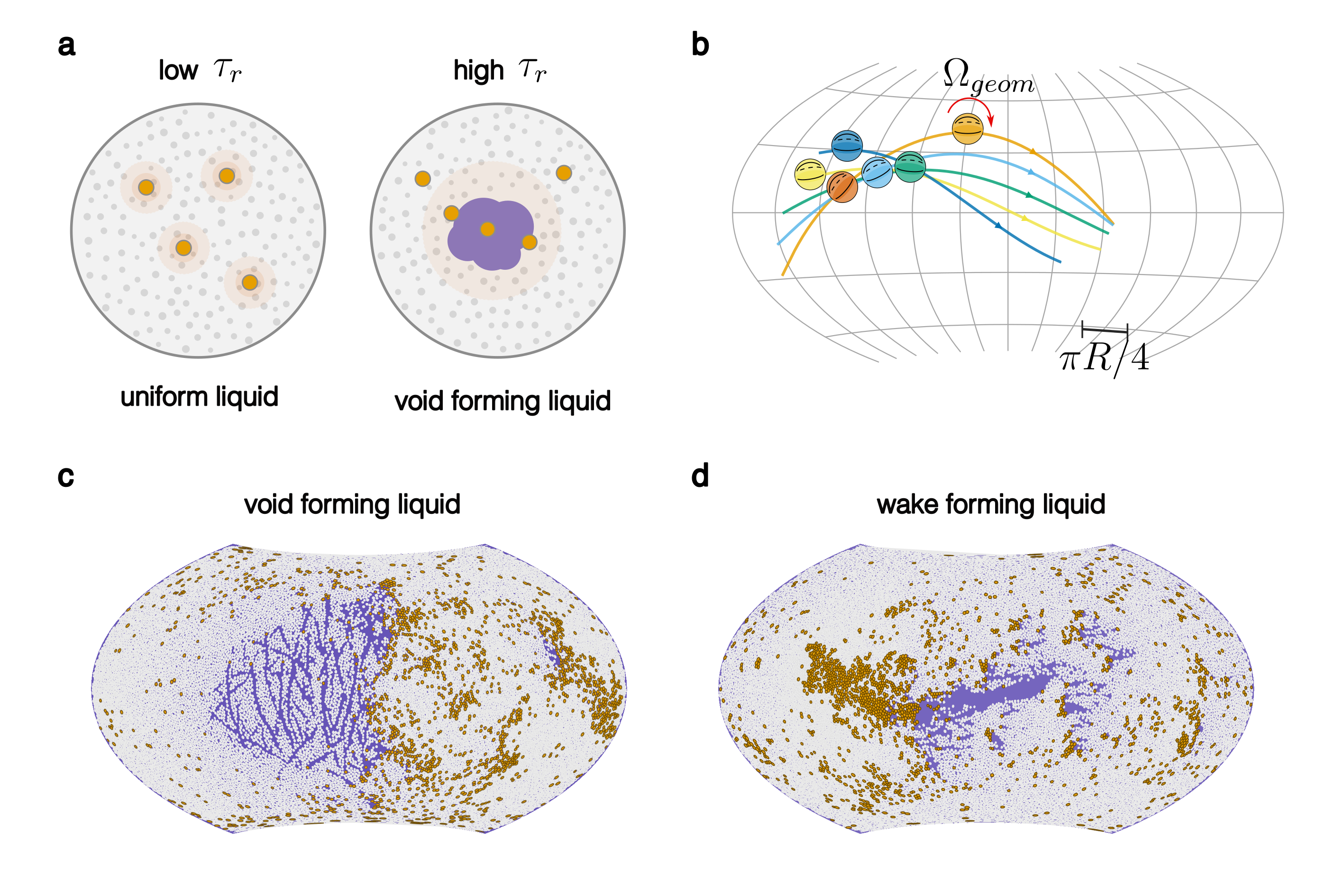}
\caption{
\textbf{Curvature and passive stress persistence organize active dopants into voids and scars.}
\textbf{a.} Schematic of the void-forming mechanism. Grey particles denote the passive host, orange particles denote active dopants, and pale purple marks passive-depleted regions. At small compactness (\(\Xi\ll1\)), passive relaxation is fast and each active dopant produces only a localized depletion cloud. At large compactness (\(\Xi\gtrsim1\)), depletion persists and the medium enters a scar-forming regime.
\textbf{b.} Schematic of geodesic transport on the sphere. A curvature-induced geometric torque, $\Omega_{geom}$, guides active particles along geodesics, so trajectories bend back through the same finite host. Particles that would drift apart on a plane can therefore re-encounter one another on the sphere and repeatedly sample the same passive material. Coloured disks denote active dopants, coloured curves denote geodesic trajectories, and the scale bar indicates a geodesic distance \(\pi R/4\).
\textbf{c.} Simulation snapshot in the void-forming regime (\(\Xi\ll1\)). Grey particles form the passive host, orange particles are active dopants, and pale purple marks the passive-depleted void. Active dopants accumulate around the depleted region.
\textbf{d.} Simulation snapshot in the wake-forming regime (\(\Xi\gtrsim1\)). Persistent passive depletion extends along the active trajectories and overlaps across geodesic distances, producing active clustering and an elongated anisotropic scar. Colors are as in \textbf{c}.}
\label{fig:1}
\end{figure*}

\begin{figure*}[t!]
\centering
\includegraphics[width=\textwidth]{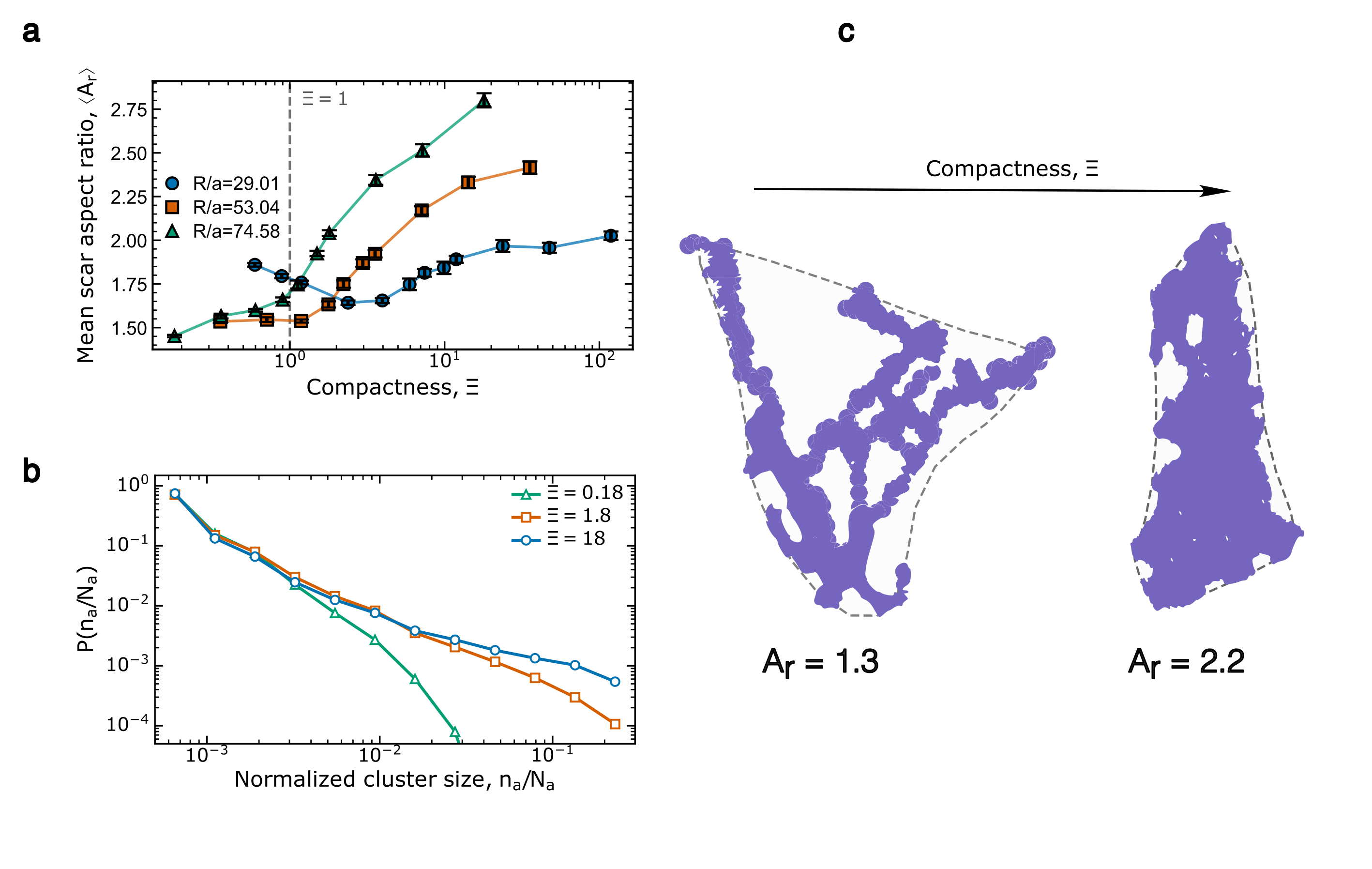}
\caption{
\textbf{Compactness controls the morphology of passive depletion.}
\textbf{a.} Aspect ratio \(A_r=PC_1/PC_2\) of the largest passive-depleted region plotted against compactness \(\Xi=(\ell_\sigma/R)^2\) for three sphere radii. The dashed line marks \(\Xi=1\), where the stress-communication length becomes comparable to the sphere radius. Across radii, \(A_r\) increases with \(\Xi\), indicating a crossover from compact passive-depleted voids to elongated scars. Symbols denote independent replicas, and error bars show the standard error of the mean (SEM) over replicas.
\textbf{b.} Active cluster-size distributions \(P(n/N_a)\) for $R/a=74.58$ and \(\Xi\approx0.18\), \(1.78\) and \(17.8\). Larger compactness shifts the distribution toward larger active clusters. Curves show the mean over independent replicas; shaded bands denote the SEM.
\textbf{c.} PCA construction used to quantify scar morphology for $R/a=74.58$. The long and short principal axes of a representative passive-depleted region define \(PC_1\) and \(PC_2\); their ratio gives the aspect ratio \(A_r\).}
\label{fig:2}
\end{figure*}
\vspace{0.5cm}
\textbf{Compactness drives the void-to-scar crossover.}
Motile or differentially diffusive particles can fluidize, restructure, or demix an otherwise crowded medium~\cite{McCarthy2024,Stenhammar2015,Weber2016}. In planar active–passive mixtures, the host response depends strongly on the persistence length of the active particles. Weakly persistent particles can fluidize a slowly relaxing passive host, whereas highly persistent particles accumulate stress in the host and nucleate passive-depleted voids~\cite{Janzen2026} (Fig.~\ref{fig:1}a). Microscopically, this depletion originates from the stress that a persistent dopant leaves in the passive host. As an active particle moves through a crowded medium, it pushes neighboring passive particles out of its path and creates an asymmetric wake behind its direction of motion. At low persistence, the host relaxes this stress before the wake grows or overlaps with others. At higher persistence, the stress perturbations generated by different dopants survive long enough to overlap, producing larger passive-depleted voids. In the plane, these voids collect and redirect active motion, but the stress field remains local and does not select a common direction.

Curvature changes this response by making active motion recurrent within the host. For active particles constrained to a smooth curved surface, persistence carries motion along geodesics, the locally straightest paths allowed by the geometry, through a curvature-induced geometric torque~\cite{Mackay2026}. On a sphere, these geodesics are great-circle arcs, so particles that would separate in the plane can re-encounter one another after traveling across the surface (Fig.~\ref{fig:1}b). In an active–passive mixture, this geometric recurrence means that dopants can repeatedly force the same host regions, allowing active-induced stress disturbances to accumulate before the passive medium relaxes.

For low persistence, the passive host relaxes before these encounters accumulate, so passive-depleted regions remain local (Fig.~\ref{fig:1}c). As persistence increases, repeated forcing of the same host regions promotes clustering inside the depleted zones. The passive response then ceases to be a set of isolated voids and becomes an extended, anisotropic depletion pattern, which we refer to as a scar (Fig.~\ref{fig:1}d). What matters is how far the host carries the stress left by a moving dopant before it relaxes. If this distance remains smaller than the sphere radius $R$, the response is local. If it becomes comparable to $R$, the same passive stress envelope spans the sphere and mechanically couples separated dopants.

We call this distance the stress-spreading length, \(\ell_\sigma\). It measures how far an active-induced stress disturbance travels through the passive host before relaxation erases it. If \(D_\sigma\) is the effective stress diffusivity of the passive contact network, estimated from contact relaxation as described in Methods, then \(\ell_\sigma=\sqrt{D_\sigma\tau_r}\).
Comparing this length with the sphere radius \(R\) defines the compactness parameter
\begin{equation}
    \Xi
    =
    \left(\frac{\ell_\sigma}{R}\right)^2
    =
    \frac{D_\sigma\tau_r}{R^2}.
    \label{eq:Xi_results}
\end{equation}
When \(\Xi\ll1\), stress explores only a small geodesic patch, so the spherical response is locally planar and active dopants nucleate compact or weakly branched passive-depleted voids. As \(\Xi\) approaches unity, the host carries stress across a distance comparable to \(R\), allowing depletion fields generated by separated dopants to overlap. The largest depleted region then elongates into a scar. Thus, at low persistence, the system behaves effectively as it does in the plane, whereas at high persistence, curvature reorganizes passive depletion into extended anisotropic structures.

This change is first visible in the shape of the depleted region. For each frame, we identify the largest depleted component and use principal component analysis to determine its two principal axes, \(PC_1\ge PC_2\). We define the aspect ratio
\begin{equation}
    A_r=\frac{PC_1}{PC_2},
\end{equation}
which measures the anisotropy of the depleted region. Compact or weakly branched voids have \(A_r=O(1)\), whereas elongated scars have \(A_r>1\). Figure~\ref{fig:2}a shows that \(A_r\) generally increases with \(\Xi\) for all three sphere radii (\(R/a=29.01\), \(53.04\) and \(74.58\), where \(a\) is the mean particle radius). The smallest sphere displays a weak nonmonotonic variation for \(\Xi<1\), which is absent for the two larger systems. We attribute this deviation to finite-size effects, since the depleted component occupies a larger fraction of the available surface and its measured principal axes are consequently more affected by fluctuations in particle positions and component shape. Radius-dependent deviations have also been reported in active–passive mixtures on spheres, where the degree of segregation changes with the ratio of sphere radius to particle size~\cite{Ai2020}. For \(\Xi>1\), the aspect ratio follows the same increasing trend observed for the larger spheres.

Although these systems reach the crossover at different microscopic relaxation times, their scar aspect ratios are organized by the same compactness parameter \(\Xi=(\ell_\sigma/R)^2\), apart from finite-size deviations for the smallest sphere at low \(\Xi\). This scaling shows that scar elongation is primarily controlled by the stress-communication length relative to the sphere radius, rather than by \(R\) or \(\tau_r\) separately. The increase in \(A_r\) marks a morphological change from compact voids, produced by local stress communication, to anisotropic scars, produced when the passive host carries stress over curvature-scale distances.

The change in host morphology also reorganizes the active dopants. Figure~\ref{fig:2}b shows the active cluster-size distribution $P(n/N_a)$ for the largest sphere, $R/a=74.58$. As compactness increases, the distribution develops a heavier tail, showing that the dopants assemble into larger clusters as scars form. The same compactness-dependent behavior is observed for the two smaller spheres, as reported in Supplementary Fig.~S1. The snapshots in Fig.~\ref{fig:1}c,d connect this aggregation to the accompanying change in host morphology. At low compactness, the depleted region remains compact and nearly isotropic, and the active particles form only small clusters. At high compactness, larger active clusters become embedded within elongated depleted regions. These scars resemble the depletion wake generated by a single dopant, but here the wakes of many dopants merge and organize collectively into an extended anisotropic structure.
\begin{figure*}[t]
    \centering
    \includegraphics[width=\textwidth]{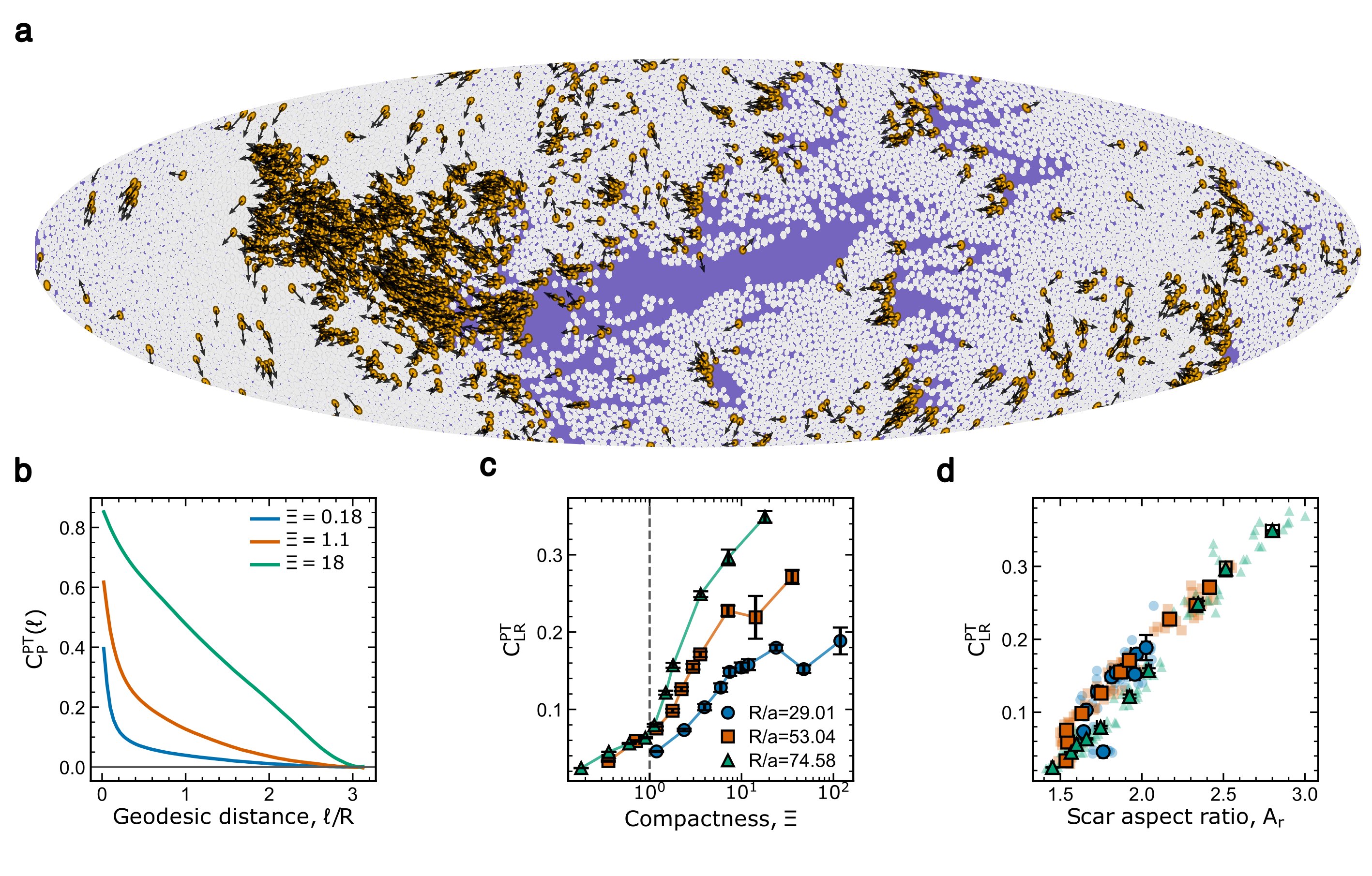}
\caption{
\textbf{Passive-scar anisotropy organizes long-range polar coherence.}
\textbf{a.} Representative snapshot for the largest sphere, \(R/a=74.58\), at high compactness (\(\Xi\approx17.8\)). Active dopants (orange disks) align along an elongated passive-depleted scar (purple region); black arrows show instantaneous orientation directors.
\textbf{b.} Parallel-transported orientational correlation \(C_P^{\rm PT}(\ell)\) versus normalized geodesic separation \(\ell/R\) for the largest sphere, \(R/a=74.58\), at \(\Xi\approx0.18\), \(1.1\) and \(18\). At low compactness, correlations decay rapidly. At high compactness, positive correlations persist to \(\ell/R=O(1)\), signaling sphere-scale polar coherence. Shaded bands denote the standard error of the mean (SEM) over independent replicas.
\textbf{c.} Long-range amplitude \(C_{\rm LR}^{\rm PT}=\langle C_P^{\rm PT}(\ell)\rangle_{1<\ell/R<2}\) as a function of compactness \(\Xi\), shown for \(R/a=29.01\), \(53.04\) and \(74.58\). The dashed line marks \(\Xi=1\). Across radii, long-range polar coherence increases as the stress-communication length becomes comparable to the sphere radius. Symbols show independent replicas, and error bars denote the SEM over replicas.
\textbf{d.} \(C_{\rm LR}^{\rm PT}\) versus scar aspect ratio \(A_r\) for the largest sphere, \(R/a=74.58\). This radius is used for the representative morphology-correlation relation in the main text; the radius dependence is shown in panel c. Long-range polar coherence increases with scar anisotropy. Faded symbols indicate independent replicas, and error bars denote the SEM across replicas.}
\label{fig:3}
\end{figure*}

\begin{figure*}[t!]
    \centering
    \includegraphics[width=\textwidth]{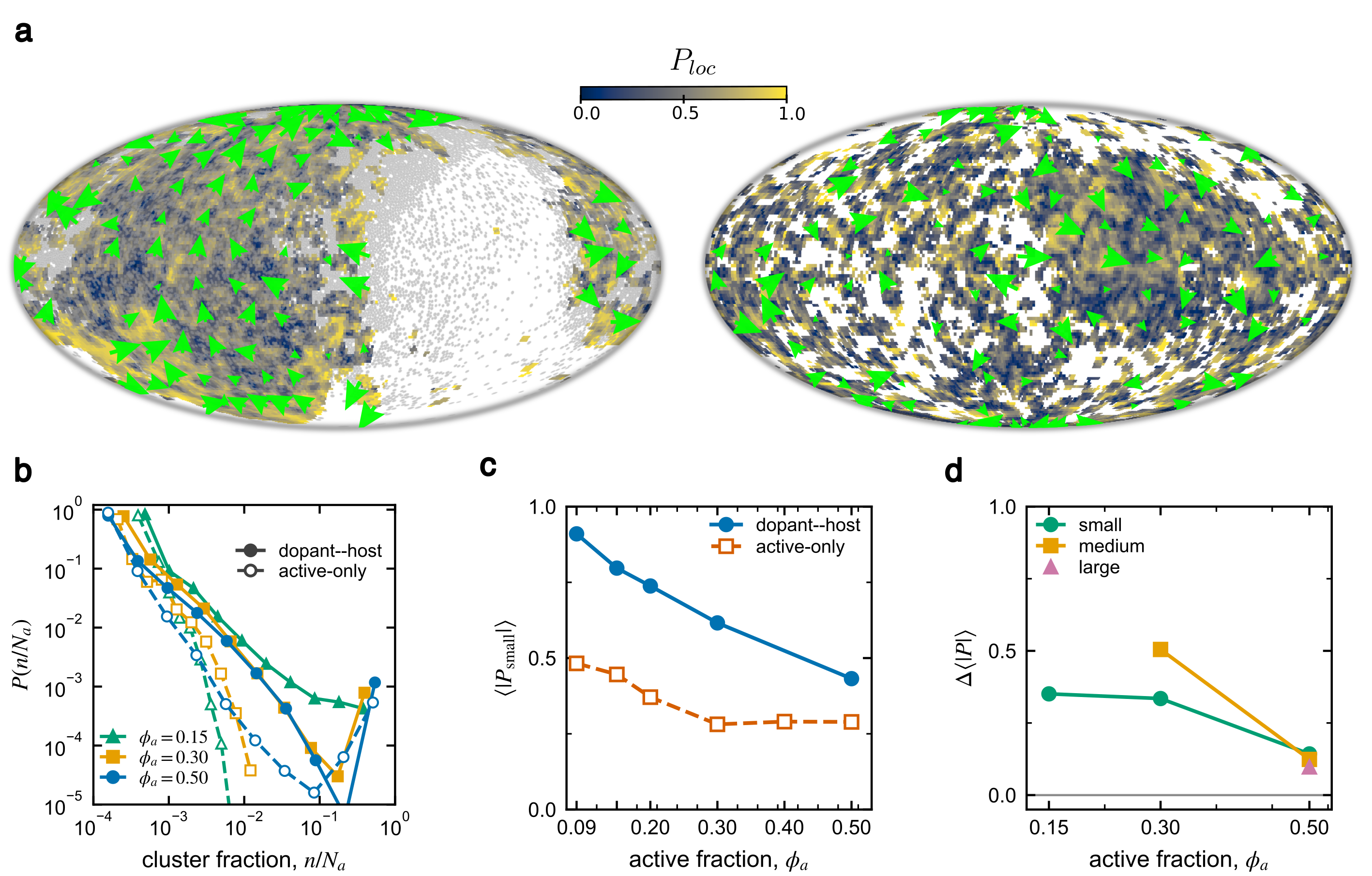}
      \caption{
\textbf{Curvature-enhanced clustering without the host does not produce polar order.}
\textbf{a.}
Mollweide projections at \(\phi_a=0.40\), comparing the coupled dopant--host system (left) with the active-only control on the same sphere (right). Background colors show the coarse-grained local polarization field \(P_{\mathrm{loc}}\), grey particles denote the passive host, and green arrows indicate the local active director field. In the coupled system, dopants form extended polar streams along passive-depleted scars. In the active-only system, the same particles aggregate but do not develop coherent polar alignment.
\textbf{b.}
Cluster-size distributions \(P(n/N_a)\) for \(\phi_a=0.15\), \(0.30\), and \(0.50\), for the host-coupled and active-only systems. The passive host enhances clustering at low and intermediate \(\phi_a\). Curves show the mean over independent replicas; shaded bands denote the SEM.
\textbf{c.}
Mean parallel-transported polarization of small clusters, \(\langle|P_{\mathrm{small}}|\rangle\), as a function of active fraction \(\phi_a\), for both systems. The host-coupled system maintains stronger cluster polarity than the active-only control across all \(\phi_a\). Error bars denote the standard error of the mean (SEM) over independent replicas.
\textbf{d.}
Host-induced excess polarization \(\Delta\langle|P|\rangle =
\langle|P|\rangle_{\mathrm{host}} -
\langle|P|\rangle_{\mathrm{only}}\)
for small, medium, and large clusters. \(\Delta\langle|P|\rangle\) decreases with increasing \(\phi_a\) and collapses once dense aggregates emerge, showing that active clustering alone does not account for the polarized state. Error bars denote the SEM over independent replicas.}
\label{fig:4}
\end{figure*}
\vspace{0.5cm}
\textbf{Passive scars organize long-range polar coherence.}
In the scar-forming regime, elongated passive depletions are accompanied by larger active clusters. Because the dopants are ABPs with no alignment rule, cluster growth alone does not imply polar order. We therefore ask whether passive scars only collect active particles or whether they also align their motion.

To quantify this coherence, we measure polar alignment between pairs of active-particle orientations. The elongated scar identifies the direction along which such alignment is expected to develop, but the pairwise comparison itself must be performed intrinsically on the sphere. Particle orientations lie in different tangent planes, so a direct dot product is not well defined. We therefore compare orientations after parallel transport along the geodesic connecting each pair and define the parallel-transported polar correlation function as
\begin{equation}
    C_P^{\rm PT}(\ell)
    =
    \bigl\langle
        \hat{\bm{p}}_i\cdot\mathcal{T}_{j\to i}\hat{\bm{p}}_j
    \bigr\rangle_{\ell_{ij}=\ell},
    \label{eq:CPT}
\end{equation}
where \(\hat{\bm{p}}_i\) is the orientation of active particle \(i\), \(\ell_{ij}\) is the geodesic separation between particles \(i\) and \(j\), and \(\mathcal{T}_{j\to i}\) parallel transports the orientation of particle \(j\) to the tangent plane of particle \(i\). This quantity is the intrinsic spherical analog of the planar polar correlation function.

To distinguish local alignment within clusters from sphere-scale coherence, we average \(C_P^{\rm PT}\) over short and long geodesic separations,
\begin{equation}
    C_{\rm SR}^{\rm PT}
    =
    \bigl\langle C_P^{\rm PT}(\ell)\bigr\rangle_{0<\ell/R<0.3},
    \qquad
    C_{\rm LR}^{\rm PT}
    =
    \bigl\langle C_P^{\rm PT}(\ell)\bigr\rangle_{1<\ell/R<2}.
    \label{eq:CLR}
\end{equation}
Here \(C_{\rm SR}^{\rm PT}\) measures polar alignment over local, cluster-scale distances, whereas \(C_{\rm LR}^{\rm PT}\) measures orientational coherence over distances comparable to the sphere radius. The short-range amplitude varies only weakly across the crossover (Supplementary Fig.~S2), confirming that local alignment and sphere-scale coherence are distinct observables.

Figure~\ref{fig:3} shows that polar order becomes long-ranged only in the scar-forming regime. At low compactness, \(C_P^{\rm PT}(\ell)\) decays rapidly with geodesic separation, indicating local orientational correlations without sphere-scale order (Fig.~\ref{fig:3}b). Equivalent systems simulated on the plane at matched density and number of dopants do not show a positive system-scale correlation tail, confirming that long-range polar coherence requires compact curved geometry rather than persistence alone (Supplementary Fig.~S3). At high compactness on the sphere, by contrast, the correlation develops a long tail that persists to \(\ell/R=O(1)\). The long-range amplitude \(C_{\rm LR}^{\rm PT}\) rises across the same compactness range in which \(A_r\) increases (Fig.~\ref{fig:3}c). Data for the different sphere radii follow the same trend when plotted against $\Xi$. Crucially, long-range polar coherence emerges for all three system sizes, with no systematic dependence of its onset on the sphere radius. This robustness distinguishes the orientational response from the finite-size deviations observed in the PCA-based scar anisotropy of the smallest sphere. The smallest sphere nevertheless exhibits greater statistical scatter, and at large $\Xi$ the correlation amplitude saturates in a radius-dependent manner, as expected when polar correlations become limited by the finite extent of the sphere. Thus, finite size affects the fluctuations and ultimate saturation of the measured correlation, but not the emergence of scar-mediated polar coherence. Polar coherence therefore grows together with the formation of anisotropic passive scars.

The line \(\Xi=1\) should therefore be understood as a crossover scale rather than a critical threshold. In the continuum stress model described in Methods, \(\Xi=(\ell_\sigma/R)^2\), with \(\ell_\sigma=\sqrt{D_\sigma\tau_r}\), compares the stress-communication length over one persistence time with the sphere radius. Thus \(\Xi\sim1\) marks the expected crossover from locally planar stress propagation to curvature-scale stress propagation. Consistent with this picture, the morphological response \(A_r\) and the orientational response \(C^{\rm PT}_{\rm LR}\) vary smoothly and increase over the same compactness range. This supports a compactness-controlled crossover from local void formation to scar-mediated polar coherence, rather than a discontinuous transition. The structural signature of that crossover is scar anisotropy.

Figure~\ref{fig:3}d shows that the long-range correlation \(C_{\rm LR}^{\rm PT}\) increases with the scar aspect ratio \(A_r\). Compact voids can collect active particles, but their nearly isotropic shape does not provide a persistent orientational bias. Elongated scars, by contrast, introduce a preferred axis that channels motion along the depleted regions. Their long-axis channels active motion along the scar and correlates the orientations of dopants separated by geodesic distances comparable to the scar length. A scar axis alone is head-tail symmetric, so channeling sets an orientation line but not a direction. The direction is fixed by the dopants themselves. The net active flux that opens the scar biases the direction of subsequent dopant motion, while the depleted channel reinforces that flux by reducing mechanical resistance along the same direction. This flux--scar feedback breaks the head--tail symmetry of the scar axis and selects polar rather than nematic coherence, consistent with the head--tail-sensitive \(C_P^{\rm PT}\) shown in Fig.~\ref{fig:3}b. The same passive structure that concentrates the dopants therefore also organizes their motion. Since the active particles have no alignment rule, this polar coherence is not imposed microscopically. It emerges from the coupling between active motion, passive stress redistribution, and compact geometry.

\vspace{0.5cm}
\textbf{Active clustering alone does not polarize.}
The results above show that passive-depleted areas (scars), active clustering, and polar coherence appear together. This coincidence leaves open the possibility that the observed polar state is a consequence of active clustering on a curved surface rather than of host-mediated stress communication. Positive curvature can increase encounters between active particles through geodesic lensing and can shift the onset of motility-induced phase separation~\cite{Fily2012,Cates2015} on curved surfaces~\cite{Schonhofer2022,Iyer2023}. Dense MIPS states can also display contact-induced velocity correlations, even without explicit alignment interactions~\cite{Caprini2020}. We therefore test whether active-only clustering on the same curved surface can reproduce the extended polar streams observed in the host-coupled system.

To test this, we compare the active--passive system with an active-only system at the same sphere radius and active fraction. The active-only system retains curvature, geodesic transport, and high-density active aggregation, but removes the passive host and hence host-mediated stress communication. We then vary the active fraction \(\phi_a\) to distinguish the scar-mediated regime from the high-density clustering expected in active-only systems.
Cluster polarization is computed intrinsically on the sphere by parallel transporting all active-particle orientations in a cluster to the tangent plane at the cluster centroid,
\begin{equation}
    |P_c|^{\rm PT}
    =
    \left|
    \frac{1}{n_c}
    \sum_{i\in c}
    \mathcal{T}_{i\to c}\hat{\bm{p}}_i
    \right|,
    \label{eq:Pc_PT}
\end{equation}
where \(\mathcal{T}_{i\to c}\) denotes parallel transport from particle \(i\) to the cluster centroid along the shortest geodesic. The same definition is used for the host-coupled and active-only systems.

Figure~\ref{fig:4}a shows the contrast at \(\phi_a=0.40\). In the presence of the passive host, active dopants form extended polar streams associated with passive-depleted scars. In the active-only case, the particles still aggregate on the same curved surface, but the aggregates remain compact and weakly polarized. Consistently, the full correlation function of the active-only control lacks the positive system-scale tail of the host-coupled state (Supplementary Fig.~S4). Thus, the passive host is not merely responsible for producing depleted scars; it is also required for the strong polar organization of the active clusters. Curvature and active clustering alone therefore do not reproduce the host-coupled state.

The active-fraction sweep makes this distinction quantitative. At low \(\phi_a\), removing the passive host strongly suppresses clustering (Fig.~\ref{fig:4}b) and reduces cluster polarization (Fig.~\ref{fig:4}c), showing that the host promotes both aggregation and orientational coherence. As \(\phi_a\) increases, the active-only system also develops larger clusters, consistent with MIPS-like aggregation~\cite{Rodriguez2020}. In the host-coupled system, the same increase in \(\phi_a\) reduces the amount of passive material available to transmit stress and sustain depleted scars. At high \(\phi_a\), the host-coupled system therefore approaches the active-only limit. Both systems form dense active aggregates, and cluster polarization decreases in both cases. The difference
\(
\Delta\langle |{\bf P}| \rangle
=
\langle |{\bf P}| \rangle_{\rm host}
-
\langle |{\bf P}| \rangle_{\rm only}
\)
measures the host-induced excess cluster polarization at fixed curvature and active fraction. Thus, \(\Delta\langle |{\bf P}| \rangle\) isolates the contribution to ordering that geodesic lensing, MIPS-like aggregation, and contact-induced velocity correlations cannot provide (Fig.~\ref{fig:4}d). Its collapse at large \(\phi_a\) shows that dense active aggregation competes with the scar-mediated state rather than generating it.

\vspace{0.5cm}
\textbf{Compact geometry enables passive-host-mediated polar order.}
Our results identify a route to polar order in which the symmetry-breaking element is generated by the environment rather than encoded in the active particles. Curvature changes how stress is communicated through the passive host. Once this communication reaches curvature-scale distances, passive depletion no longer remains a collection of local voids. It forms anisotropic scars that organize otherwise non-aligning dopants into polar streams.

This mechanism is distinct from two simpler alternatives. The first is geometric confinement by a void boundary. In planar active--passive mixtures, long-lived voids also trap and redirect dopants, but their boundaries remain compact or branched and no coherent polarity emerges~\cite{Janzen2026}. A void can collect motion without polarizing it. Polarity appears only when the passive-depletion field becomes anisotropic and system-spanning. The second alternative is curvature-enhanced MIPS. This is a stringent control because curvature alone can change both active aggregation and cluster morphology. Positive curvature can act as a geometrical lens and shift the MIPS threshold of active-only particles~\cite{Schonhofer2022}, while spherical confinement introduces the time scale \(R/v_0\) and reshapes the MIPS phase diagram of ABPs~\cite{Iyer2023}. More generally, curved geometry can control where dense MIPS clusters form and whether they appear as compact domains or extended bands~\cite{webb2026}. Dense active phases can also generate velocity correlations without an explicit alignment rule~\cite{Caprini2020,Caprini2023}. Our active-only control retains these ingredients. It keeps the same sphere, active fraction and contact-induced reorientation, but removes the passive host and therefore the passive stress field. The particles still aggregate, yet they do not form scar-guided polar streams. The relevant distinction is therefore not clustering versus no clustering. It is an internally disordered active aggregate versus a host-coupled scar state whose polarity is selected by an anisotropic passive depletion field.

\textbf{Conclusions.}
The mechanism identified here is scar-mediated polarity. Compact geometry lets a passive host transmit active-induced stress over distances comparable to the system size. When \(\Xi\ll1\), stress communication stays local and active dopants form compact passive-depleted voids. When \(\Xi\gtrsim1\), the same stress field spans the sphere, elongates the depleted region into a scar and gives non-aligning dopants a common direction of motion. The symmetry-breaking element is set by the environment, not by an alignment rule in the active particles. The active-only comparison supports this interpretation. At matched curvature and active fraction, the active-only system still clusters, yet it does not develop comparable long-range polar order, which separates scar-mediated polarity from motility-induced phase separation~\cite{Schonhofer2022,Iyer2023}.

In soft-matter systems, scar-mediated polarity offers a geometric means of controlling collective motion without modifying the active particles themselves. At fixed activity and composition, changing the system radius tunes the ratio $\ell_\sigma/R$ and can therefore drive the system across the scar-forming crossover. From $\Xi=D_\sigma\tau_r/R^2$, the crossover condition $\Xi\sim1$ predicts that the characteristic persistence time should scale as $\tau_r^{c}\sim R^2/D_\sigma$. Synthetic active materials provide a natural setting in which to test this scaling because activity, confinement, and droplet size can be tuned independently~\cite{Birrer2022,Kato2025,Keber2014}. The expected signature is the emergence of long-range parallel-transported polar correlations, $C^{\rm PT}_{\rm LR}$, when $\ell_\sigma$ becomes comparable to $R$. These correlations should grow together with the scar aspect ratio $A_r$, rather than simply with active cluster size, and should disappear when the passive host is removed or sufficiently diluted, even if active aggregation persists.

The same mechanism may extend to biological active materials, where the passive host is replaced by a deformable and continuously rearranging environment. Epithelial monolayers on curved substrates provide a natural setting, since curvature can be imposed while cell shape and collective motion are measured~\cite{Pieuchot2018,Glentis2022}. Whereas confining geometry can organize such tissues through surface-anchored equilibrium ordering~\cite{Guruciaga2026}, the polarity here is selected dynamically by active stress redistribution through the host. A key test would be whether collective polarity emerges together with an anisotropic deformation or stress pattern in the surrounding tissue, rather than with cell clustering alone.

More broadly, the compactness mechanism is not specific to the sphere. On any compact surface, the passive response crosses over from locally planar to geometry-spanning when the stress-communication length becomes comparable to the longest accessible geodesic scale. For a general compact surface \(\mathcal{M}\), this scale is set by the lowest non-zero eigenvalue \(\lambda_1(\mathcal{M})\) of the Laplace--Beltrami operator, giving the spectral criterion
\(
D_\sigma \tau_r \lambda_1(\mathcal{M}) \sim 1.
\) The spherical compactness parameter \(\Xi=D_\sigma\tau_r/R^2\) is therefore one realization of a more general spectral criterion. Surfaces with several well-separated low-lying modes may exhibit successive crossovers as stress communication reaches different geometric scales, while topology constrains the polar textures that can be supported globally. Scar-mediated polarity thus provides a general route by which compact geometry and passive stress transmission can organize non-aligning active particles.

\section*{Methods}

\textbf{Simulation model and geometry.}
We study $N$ polydisperse disks confined to the surface $\mathbb{S}^2(R)$ of a sphere of radius $R$, extending to curved geometry the planar active--passive mixture of Ref.~\cite{Janzen2026}. Radii $a_i$ are drawn from a uniform distribution with mean $a$ and half-width $0.2\sqrt{3}\,a$ (20\% polydispersity, sufficient to suppress crystallization). Particles interact through purely repulsive harmonic contact forces,
\begin{equation}
    \mathbf{F}_{ij} = k\left(a_i + a_j - r_{ij}\right)\hat{\mathbf{r}}_{ij}
    \quad \text{for } r_{ij} < a_i + a_j,
\end{equation}
and $\mathbf{F}_{ij}=\mathbf{0}$ otherwise, where $r_{ij}$ is the geodesic distance between particles $i$ and $j$, $\hat{\mathbf{r}}_{ij}$ is the unit tangent vector along the geodesic connecting them, and $k$ is the contact stiffness. The main system has $R=74.58\,a$ and $N\approx2\times10^4$ particles at packing fraction $\phi=0.9$; finite-size results are reported below.

\textbf{Active and passive particles.}
A fraction $\phi_a=N_a/N$ of particles are active Brownian particles (ABPs). Each ABP self-propels along an instantaneous director $\mathbf{p}_i$, a unit tangent vector on the sphere, with speed $v_0$. The remaining $N-N_a$ particles are passive and are thermalized by translational Brownian noise alone. The overdamped equations of motion for particle positions $\mathbf{r}_i$ on $\mathbb{S}^2(R)$ are
\begin{equation}
    \dot{\mathbf{r}}_i = \mu\sum_{j\neq i}\mathbf{F}_{ij}
  + v_0\,\mathbf{p}_i\,\delta_{i\in\mathrm{active}}
    + \sqrt{2D_t}\,\boldsymbol{\eta}_i,
\end{equation}
where $\mu$ is the mobility, $D_t$ the translational diffusivity of the passive particles, and $\boldsymbol{\eta}_i$ is unit-variance Gaussian white noise projected onto the local tangent plane. Positions are constrained to the sphere after each update. The director of each ABP undergoes geodesic parallel transport along its trajectory and is subject to rotational diffusion in the tangent plane,
\begin{equation}
    \dot{\mathbf{p}}_i = \sqrt{2D_r}\,\xi_i\,\hat{\mathbf{e}}_\perp(\mathbf{p}_i),
\end{equation}
where $\xi_i$ is scalar Gaussian white noise and $\hat{\mathbf{e}}_\perp(\mathbf{p}_i)$ is the unit tangent vector perpendicular to $\mathbf{p}_i$ in the local tangent plane. This defines the persistence time $\tau_r=D_r^{-1}$ and persistence length $\ell_p=v_0\tau_r$. Active--passive coupling enters only through the repulsive contact force; there is no explicit alignment rule.

\textbf{Control parameters and compactness.}
Lengths are measured in units of the mean radius $a$ and times in units of the elastic relaxation time $\tau_0=(\mu k)^{-1}$. The dimensionless control parameters are $\tilde{v}=v_0/(a\mu k)$ and $\tilde{D}_r=D_r/(\mu k)$. We fix $\tilde{v}=0.1$ and vary $\tilde{D}_r$ to sample persistence times $\tau_r\in\{10^3,10^4,10^5\}\,\tau_0$. Passive particles have $\tilde{D}_t=D_t(\mu k)/a^2=10^{-7}$, placing them deep in the mechanically arrested regime in the absence of active forcing. The active fraction is fixed at $\phi_a=0.09$ unless stated otherwise.

The effective stress diffusivity $D_\sigma$ is estimated from the microscopic contact dynamics. An elastic contact of stiffness $k$ relaxes on time scale $\tau_0=(\mu k)^{-1}$ since $\dot{x}=-\mu k x$. If one relaxation event redistributes stress over a particle-scale distance $a$, then $D_\sigma\sim a^2/\tau_0=\mu ka^2$, consistent with the stress-transmission picture of dense soft-particle and amorphous materials~\cite{Nicolas2018,Bocquet2009,Goyon2010,Ness2022}. The stress-spreading length over one persistence time is $\ell_\sigma=\sqrt{D_\sigma\tau_r}\sim a\sqrt{\tau_r/\tau_0}$, giving the compactness parameter
\begin{equation}
    \Xi = \left(\frac{\ell_\sigma}{R}\right)^2 \sim \frac{\tau_r/\tau_0}{(R/a)^2}.
    \label{eq:xi_estimate}
\end{equation}
For $R=74.58\,a$, the three simulated persistence times correspond to $\Xi\approx0.18$, $1.78$ and $17.8$. The onset of scar formation visible in Fig.~\ref{fig:2} occurs at $\tau_r\approx6\times10^3\,\tau_0$, giving $\Xi\approx1.07$, consistent with the theoretical crossover $\Xi\sim1$.
Finite-size effects were assessed at two additional sphere radii, $R/a=53.04$ and $29.01$. At matched $\Xi$, the $R/a=53.04$ system shows the same scar-forming phenomenology as the main system, indicating that the results are not sensitive to particle number at fixed $\phi$. The smallest sphere, $R/a=29.01$, already lies at $\Xi\approx1.19$ for the shortest simulated persistence time, $\tau_r=10^3\,\tau_0$, and therefore does not access the regime $\Xi\ll1$. This follows directly from the scaling $\Xi\sim(\tau_r/\tau_0)/(R/a)^2$ and provides an independent consistency check of the estimate $D_\sigma\approx a^2/\tau_0$.

\textbf{Integration and sampling.}
Equations of motion are integrated with a forward-Euler scheme at time step $dt=5\times10^{-3}$ using the GPU-accelerated \textsc{SAMoS} package~\cite{Sknepnek2024}. Each run comprises an equilibration phase of $5\times10^6\,\tau_0$ followed by a production phase of equal duration; all reported observables are computed from production trajectories. At the largest persistence time ($\tau_r=10^5\,\tau_0$), the production window covers $50\,\tau_r$, sufficient for scar morphology and orientational statistics to reach stationary values (Supplementary Fig.~S5). Results are averaged over $N_{\rm runs}=10$ independent realizations initialized from distinct quenched configurations.

\textbf{Measured observables.}
\textit{Scar morphology.} Passive-depleted regions are identified from the coarse-grained passive density on the sphere. For each production frame, the largest depleted component is extracted and mapped to its local tangent plane; principal component analysis gives axes $PC_1\ge PC_2$ and the scar aspect ratio $A_r=PC_1/PC_2$.

\textit{Cluster-size distribution.} Active particles are grouped into clusters by contact connectivity. The distribution $P(n/N_a)$ gives the fraction of active particles belonging to clusters of size $n$, normalized by the total active-particle count $N_a$.

\textit{Orientational correlations.} The geodesic separation between two active particles at positions $\bm{r}_i$ and $\bm{r}_j$ is
\begin{equation}
    \ell_{ij}
    = R\arccos\!\bigl(\hat{\bm{r}}_i\cdot\hat{\bm{r}}_j\bigr),
    \qquad
    \hat{\bm{r}}_k=\bm{r}_k/R.
    \label{eq:geodesic_dist}
\end{equation}
The parallel-transported orientational correlation $C_P^{\rm PT}(\ell)$ (Eq.~\eqref{eq:CPT}) and its short- and long-range amplitudes $C_{\rm SR}^{\rm PT}$ and $C_{\rm LR}^{\rm PT}$ (Eq.~\eqref{eq:CLR}) are pair-count-weighted averages over geodesic bins $0<\ell/R<0.3$ and $1<\ell/R<2$ respectively.

\textit{Cluster polarization.} The parallel-transported cluster polarization $|P_c|^{\rm PT}$ (Eq.~\eqref{eq:Pc_PT}) is the magnitude of the mean active-particle director in a cluster, with all directors transported to the cluster centroid along the shortest geodesic. Small clusters are defined as those with $n/N_a < 0.01$; clusters with $0.01 \le n/N_a < 0.05$ are classified as medium and those with $n/N_a \ge 0.05$ as large, corresponding to the three bins in Fig.~\ref{fig:4}d. The quantity $\langle|P_{\rm small}|\rangle$ is obtained by averaging the cluster polarization over all small clusters within each snapshot, then averaging the snapshot-level values over the production frames of each replica before computing the replica mean and SEM.

\textbf{Replica averaging and SEM.}
For each observable, production-frame values are first averaged within each independent realization to produce one replica-level estimate. The reported mean is the arithmetic mean over $N_{\rm runs}=10$ replicas. Error bars and shaded bands report the standard error of the mean,
\[
    \mathrm{SEM} = s/\sqrt{N_{\rm runs}},
\]
where $s$ is the sample standard deviation across replicas. For distributional observables, histograms are normalized within each replica before computing the replica mean and SEM bin by bin.

\textbf{Stress-diffusion theory on \(\mathbb{S}^2(R)\).}
The coarse-grained mechanical response of the passive host is described by a scalar stress field $\sigma(\mathbf{x},t)$ on $\mathbb{S}^2(R)$,
\begin{equation}
\partial_t\sigma
=
D_\sigma \,\Delta_{\mathrm{LB}} \sigma
-
\frac{\sigma}{\tau_m}
+
s(\mathbf{x},t),
\label{eq:PDE_main}
\end{equation}
where $D_\sigma$ is the stress diffusivity, $\tau_m$ the passive relaxation time, $\Delta_{\mathrm{LB}}$ the Laplace--Beltrami operator, and $s$ the active forcing density. Here $\sigma$ is a scalar invariant of the local stress tensor, such as the coarse-grained pressure, so the model captures the geodesic range of mechanical stress communication rather than the full tensorial elastic response. The directional character of scar formation is not encoded in $\sigma$ itself, it arises because a persistently moving dopant generates a stress perturbation that is asymmetric along its trajectory. The scalar model controls how far this perturbation spreads across the sphere; the scar orientation is set by the net active flux direction, which the passive depletion then reinforces. This is the curved-space extension of the planar stress-diffusion model of Refs.~\cite{MatozFernandez2017,Nicolas2018,Janzen2026}.

The Green's function $G(\gamma,t)$, where $\gamma$ is the geodesic angle between source and field point, follows from the spherical-harmonic decomposition of Eq.~\eqref{eq:PDE_main},
\begin{equation}
G(\gamma,t)
=
\frac{e^{-t/\tau_m}}{4\pi R^2}
\sum_{\ell=0}^{\infty}
(2\ell+1)\,
\exp\!\left(-D_\sigma t\,\frac{\ell(\ell+1)}{R^2}\right)
P_\ell(\!\cos\gamma),
\label{eq:G_exact}
\end{equation}
where $P_\ell$ are Legendre polynomials and the eigenvalues of $-\Delta_{\mathrm{LB}}$ are $\lambda_\ell=\ell(\ell+1)/R^2$. For $D_\sigma t\ll R^2$ and $\gamma\ll1$, a saddle-point evaluation using the Mehler--Heine approximation $P_\ell(\cos\gamma)\approx J_0((\ell+\tfrac12)\gamma)$ and the Hankel identity $\int_0^\infty q\,e^{-aq^2}J_0(qr)\,dq=(2a)^{-1}e^{-r^2/4a}$ recovers the planar Gaussian,
\begin{equation}
G(\gamma,t)
\approx
\frac{e^{-t/\tau_m}}{4\pi D_\sigma t}
\exp\!\left(-\frac{r^2}{4D_\sigma t}\right),
\qquad r = R\gamma,
\label{eq:G_local}
\end{equation}
with leading curvature correction $[1+Kr^2/12+KD_\sigma t/3+\cdots]$, where $K=1/R^2$.

The mechanically relevant quantity is the stress dose accumulated during one persistence interval. For $\tau_r\lesssim\tau_m$,
\begin{equation}
I(\gamma)
=
\sigma_0\int_0^{\tau_r}(\tau_r - u)\,G(\gamma,u)\,du.
\label{eq:dose_def}
\end{equation}
Inserting Eq.~\eqref{eq:G_exact} and evaluating the time integral in closed form yields
\begin{equation}
\begin{split}
I(\gamma)
&\approx
\frac{\sigma_0}{4\pi R^2}
\sum_{\ell=0}^{\infty}
(2\ell+1)\,P_\ell(\!\cos\gamma)\\
&\quad\times
\left[
\frac{\tau_r}{\alpha_\ell}
-
\frac{1-e^{-\alpha_\ell\tau_r}}{\alpha_\ell^2}
\right],\\
&\alpha_\ell = D_\sigma\frac{\ell(\ell+1)}{R^2}.
\end{split}
\label{eq:dose_spectral}
\end{equation}
with the $\ell=0$ term, obtained by taking $\alpha_0\to0$, giving the spatially uniform background dose $I_0=\sigma_0\tau_r^2/(8\pi R^2)$. The passive medium locally yields where $I(\gamma)\ge\Sigma_y$, defining the single-dopant influence region as a spherical cap of area $V_{\rm infl}=2\pi R^2(1-\cos\gamma_{\rm infl})$.

The crossover between local and system-scale stress communication is governed by
\begin{equation}
\Xi = \left(\frac{\ell_\sigma}{R}\right)^2,
\qquad
\ell_\sigma = \sqrt{D_\sigma\tau_r}.
\label{eq:Xi_def}
\end{equation}
For $\Xi\ll1$, high-$\ell$ modes control the kernel, $G$ reduces to the planar Gaussian (Eq.~\eqref{eq:G_local}), and the influence region is a compact disk of radius $\sim\ell_\sigma$. For $\Xi\gtrsim1$, the low-$\ell$ modes, including the uniform $\ell=0$ contribution, dominate the integrated dose, and influence regions from separated dopants overlap across geodesic distances.

On any compact surface $M$, the stress Green's function takes the spectral form
\begin{equation}
G(x,y;t)
=
e^{-t/\tau_m}
\sum_{n=0}^{\infty}
e^{-D_\sigma\lambda_n t}
\psi_n(x)\psi_n^*(y),
\label{eq:G_general_compact}
\end{equation}
where $\lambda_n$ and $\psi_n$ are eigenvalues and eigenfunctions of $-\Delta_{\rm LB}$~\cite{Grigoryan2009}. At short persistence times, high modes govern the local kernel and the response is locally planar. At longer times, modes with $D_\sigma\tau_r\lambda_n\gg1$ are filtered out and the lowest non-uniform modes set the geometry of stress communication. The spherical criterion $\Xi\sim1$ is therefore the $\mathbb{S}^2(R)$ realization of the general spectral condition
\begin{equation}
D_\sigma\tau_r\,\lambda_1(M)\sim1,
\label{eq:spectral_compactness}
\end{equation}
up to factors of order unity. Surfaces with several geometric lengths produce sequential crossovers as different modes are filtered out. Non-uniform curvature may localize the first compact stress structures in high-curvature regions.

\section*{Data availability}
The source data underlying the figures are provided with this paper. Additional data generated or analysed during this study are available from the corresponding authors upon reasonable request.

\section*{Acknowledgements}
D.A.M.-F. and G.J. acknowledge support from the Spanish Ministry of Science, Innovation and Universities and the Agencia Estatal de Investigaci\'on through grant PID2023-148991NA-I00. D.A.M.-F. also acknowledges support from the Ram\'ony Cajal Programme through grant RYC2023-043002-I.

\section*{Author contributions}
G.J.\ and D.A.M.-F.\ contributed equally to all aspects of this work.

\section*{Competing interests}
The authors declare no competing interests.
\bibliographystyle{apsrev4-1} 
\bibliography{references}
\clearpage
\end{document}